Rozdział 6

# ZASTOSOWANIE GRAFÓW I SIECI W SYSTEMACH REKOMENDACJI

Michał Malinowski[17]

**Streszczenie:** Celem niniejszego rozdziału jest przedstawienie wykorzystania teorii grafów i sieci w obszarze rekomendacji, w tym modeli matematycznych stanowiących podstawę do konstruowanych na ich bazie algorytmów i systemów rekomendacji. W pierwszej części rozdziału został zawarty syntetyczny opis obszaru rekomendacji, ze szczególnym uwzględnieniem typów rozwiązań rekomendacyjnych oraz matematycznego opisu problemu. W kolejnej części opracowania zostały przedstawione modele i techniki wykorzystania grafów i sieci oraz przykładowe algorytmy zbudowane na ich bazie.

**Słowa kluczowe:** system rekomendacji, algorytm rekomendacji, grafy i sieci

## Wprowadzenie

Rozwiązania z obszaru rekomendacji stają się coraz bardziej istotne w nowoczesnej gospodarce konsumenckiej i społeczeństwie informacyjnym. Systemy tego typu są wykorzystywane m.in. w dużych portalach internetowych, wyszukiwarkach sieciowych, serwisach społecznościowych oraz sklepach online. Otoczenie informacyjne, w którym żyjemy, a zwłaszcza ogromna ilość przekazywanych informacji uniemożliwia efektywne ich wykorzystanie (Malinowski, Sokólski 2001). W tym celu tworzone są dedykowane algorytmy, a co za tym idzie, oparte na nich systemy informatyczne, których główną funkcją jest wsparcie w procesie podejmowania decyzji w natłoku informacji.

Dlaczego „systemy rekomendacji", a nie „systemy rekomendacyjne"? Oba pojęcia funkcjonują w literaturze równolegle, jednak bardziej popularnym określeniem jest „system rekomendacji". Za miarę popularności służą wyniki wyszukiwarki Google, które są następujące[18]:
− system rekomendacji – około 13 000 tys. wyników;
− system rekomendacyjny – około 24,8 tys. wyników.

## Obszar badawczy

Obszar rekomendacji rozpatrywany jest w trzech głównych płaszczyznach (Liling 2019):

---

[17] Wojskowa Akademia Techniczna, Wydział Cybernetyki
[18] Pomiar przeprowadzony w dniu 27 sierpnia 2021 r. na bazie wyszukiwarki internetowej Google (https://google.pl)

– technik rekomendacji – w kontekście rozwoju metod analitycznych problematyki rekomendacji;
– algorytmów rekomendacji – w kontekście opracowywania algorytmów opartych na technikach rekomendacji;
– systemów rekomendacji – w kontekście budowy rozwiązań informatycznych wspierających procesy rekomendacji wykorzystujących algorytmy rekomendacji.

Systemy rekomendacji służą do oszacowania preferencji użytkowników dotyczących przedmiotów, usług lub obiektów, których jeszcze nie widzieli lub nie znają. Systemy rekomendacji często używają danych wejściowych, takich jak preferencje użytkownika, cechy (atrybuty) przedmiotu (obiektu), historia przeszłych interakcji użytkowników z przedmiotami (obiektami), dane czasowe i dane przestrzenne (Bhaskar, Raja 2020).

Rekomendacje dostarczane przez tego typu systemy mają dwie różne formy:
– Przewidywanie ocen – w tym przypadku szacowana jest ocena użytkownika dla nowego przedmiotu. Przewidywanie bazuje na podstawie ocen rozpatrywanego obiektu dokonanych przez innych użytkowników. Dla przykładu przewidywaniem oceny jest prognoza odpowiadająca na pytanie: „Czy konkretnemu użytkownikowi spodoba się określony film?". W serwisie Netflix prognoza tego typu bazuje na podstawie historycznego zachowania innych użytkowników tego serwisu (Symeonidis i in. 2011).
– Prognoza rankingu – szacowanie wyniku polega na utworzeniu listy rankingowej „Top-N" z N pozycjami dla konkretnego użytkownika. System bazujący na tym schemacie może polecać „10 najlepszych książek do przeczytania, jeśli lubisz Harry'ego Pottera" (Cremonesi i in. 2010).

W rozwiązaniach e-Commerce preferowane są systemy rekomendacji dające wyniki w postaci „prognoz rankingowych", a nie „przewidywanych ocen", ponieważ firmy wolą wiedzę na temat oczekiwanych (preferowanych) produktów niż ich ocenę (Steck 2013).

Najważniejszą cechą systemu rekomendacji jest zdolność do przewidywania preferencji i zainteresowań użytkownika poprzez analizę jego zachowania lub zachowania innych użytkowników, w celu wygenerowania spersonalizowanych wyników. Aktualnie w opracowaniach dotyczących obszaru rekomendacji wyróżnia się następujące główne typy technik rekomendacji (Ricci i in. 2015, s. 10):
– *Content Based Methods*,
– *Collaborative Filtering Methods*,
– *Memory-Based*,
– *Model-Based*,
– *Hybrid Methods*.

## *Content Based Methods*

Algorytmy bazujące na tej technice są konstruowane na bazie założenia, że obiekty o podobnych funkcjach otrzymają podobne oceny od tego samego użytkownika. W konsekwencji systemy tego typu polecają obiekty, które są podobne do przedmiotów, w stosunku do których użytkownik wykazywał zainteresowanie w przeszłości. Podobieństwo obiektów jest obliczane na podstawie określonych atrybutów i z wykorzystaniem różnych metod (Li, Cai, Liao 2012).

## *Collaborative Filtering Methods*

Idea tej techniki opiera się na założeniu, że jeśli dwóch użytkowników podejmowało w przeszłości podobne decyzje, to w przyszłości ich preferencje też będą zbieżne. Przykładem może być sklep z grami planszowymi, w którym dwaj użytkownicy wykazali zainteresowanie podobnymi grami, ale jeden z nich zainteresował się jeszcze innymi grami. W takiej sytuacji mechanizm wspólnej filtracji zarekomenduje te gry również drugiemu użytkownikowi, zakładając, że i jemu też przypadną one do gustu. W powyższej sytuacji system nie opiera swojego działania na atrybutach obiektów, a jedynie na zachowaniu osób, których zainteresowania i preferencje zostały uznane za podobne (Su, Khoshgoftaar 2009).

## *Memory-Based*

Charakteryzują się one tym, że do przeprowadzenia rekomendacji używają całej bazy danych serwisu. Zakładają one, że każdy użytkownik jest częścią pewnej grupy, dlatego używają metod statystycznych do znalezienia tzw. „najbliższych sąsiadów", czyli grupy użytkowników o podobnych zainteresowaniach. W pierwszym kroku mechanizm rekomendacji tego typu wyszukuje grupę najbliższych sąsiadów poprzez obliczenie wagi $w_{i,j}$, która określa podobieństwo lub korelację pomiędzy dwoma użytkownikami: $i$ oraz $j$. Następnie obliczana jest prognoza określająca, czy dany produkt może być polecony użytkownikowi w oparciu o jego najbliższych sąsiadów. Ostatnim krokiem jest wybór $N$ obiektów najbardziej pasujących do użytkownika i przedstawienie ich w postaci rekomendacji (Su, Khoshgoftaar 2009).

## *Model-Based*

Podstawą ich działania jest utworzenie modelu ocen użytkownika, który będzie w stanie przewidzieć jego ocenę dotyczącą obiektów. Podczas tworzenia modelu wykorzystywane są techniki uczenia maszynowego na podstawie danych treningowych (Sarwar i in. 2001).

Jednym ze sposobów oszacowania rekomendacji obiektu dla danego użytkownika jest użycie klasyfikatora bayesowskiego, który jest najczęściej stosowany w systemach rekomendacyjnych. Stanowi on jedną z metod uczenia maszynowego, określającą, do której z klas decyzyjnych należy przypisać nowy przypadek. W kontekście mechanizmów rekomendacji określane jest, jak bardzo wybrany obiekt jest odpowiedni dla danego użytkownika (Miyahara, Pazzani 2000).

## *Hybrid Methods*

Techniki tego typu wykorzystują właściwości obu wcześniej opisanych podejść i stanowią kompilację zarówno techniki *Content Based Filtering Methods*, jak i *Collaborative Filtering* (CF) *Based Methods*. Umożliwiają eliminowanie wad poszczególnych podejść. Techniki hybrydowe w dużej mierze wykorzystują profile użytkowników i opisy obiektów, aby znaleźć użytkowników o podobnych zainteresowaniach, a następnie wykorzystują filtrowania oparte na współpracy w celu prognozowania (Cremonesi i in. 2010).

## Problem badawczy

W oparciu o techniki, które funkcjonują i są rozwijane w sferze idei, opracowywane są algorytmy rekomendacji. W efekcie różne algorytmy mogą być opracowywane na podstawie tych samych technik i dawać te same wyniki. Istotnymi różnicami pomiędzy nimi są czasy generowania wyników oraz wielkości potrzebnych zasobów (moc procesora i pamięć) (Goodrich, Tamassia 2001, "2.2" section). Docelowo bowiem algorytmy są implementowane z wykorzystaniem technik komputerowych w systemach rekomendacji, które stosowane są w ramach rozwiązań e-Commerce.

Generalnie zadaniem algorytmów rekomendacji jest rozwiązanie zadania rekomendacyjnego opisywanego w następujący sposób (Esmaili i in. 2006). Niech $C$ będzie zbiorem użytkowników wykorzystujących system rekomendacji i $O$ zbiorem wszystkich możliwych obiektów, które mogą być rekomendowane. Ponadto niech $u$ będzie funkcją użyteczności mierzącą użyteczność obiektu $o \in O$ dla użytkownika $c \in C$. To jest $u: C x O \to W$, gdzie $W$ jest uporządkowanym zbiorem (np. nieujemne liczby całkowite). Następnie dla każdego użytkownika $c$ należącego do $C$ zostaje wybrany podzbiór $R_c^*$ należący do $O$, który maksymalizuje użyteczność dla użytkownika. Oznacza to, że dla ustalonego $c \in C$ wyznaczamy zbiór rekomendowanych obiektów następująco:

$$R_c^* = \{r \in O: u(c,r) = \max_{o \in O} u(c,o) \} \tag{1}$$

W przypadku ogólnym zadaniem algorytmów rekomendacyjnych jest znalezienie podzbioru $R_c^*$, zwanego zbiorem rekomendowanych obiektów dla użytkownika $c$.

Istnieją dwa główne problemy w budowie takich algorytmów: obsługa nielicznych danych użytkowników (Shams, Haratizadeh 2017) oraz potrzeba niejawnie wyprowadzonych informacji (McAuley, Leskovec 2013). W związku z powyższym zadanie ulega modyfikacji polegającej na tym, że rekomendacji dokonuje się względem wybranego obiektu $m$ bez bezpośredniego uwzględnienia użytkownika $c$, dla którego rekomendacja następuje. W praktyce takie zdarzenie zachodzi wówczas, gdy system rekomendacyjny nie posiada informacji na temat użytkownika. Może mieć to miejsce w następujących sytuacjach: użytkownik po raz pierwszy ma kontakt z systemem lub system nie gromadzi informacji o użytkownikach.

W związku z powyższym zadanie (1), uwzględniając zmianę w funkcji użyteczności taką, że $u: O x O \rightarrow W$ ulega zmianie tak, że zbiór rekomendacji dla użytkownika $R_c^*$ zostaje zastąpiony przez $R_m^*$, czyli zbiór rekomendacji dla obiektu. W konsekwencji, dla ustalonego $m \in O$, zostaje wyznaczony zbiór rekomendowanych obiektów następująco:

$$R_m^* = \{r \in O: u(m,r) = \max_{o \in O} u(m,o) \} \qquad (2)$$

Uwzględnienie użytkownika $c$ polega na tym, że to on, świadomie kierując swoim zachowaniem, dokonuje wyboru obiektu $m$ ze zbioru $O$, względem którego będzie budowana rekomendacja $R_m^*$.

## Implementacja struktur grafowych

Jednym z głównych problemów przy implementacji algorytmów rekomendacji bazujących na grafach i sieciach, zwanych *Graph-Based Recommender System*, określanych skrótem GRS (Hekmatfar, Haratizadeh, Goliaei 2021) w systemach rekomendacji, jest odwzorowanie struktury grafu w strukturze pamięci komputera lub bazie danych docelowego systemu. Z punktu widzenia informatyki naturalnym rozwiązaniem w tej sytuacji wydaje się zastosowanie bazy grafowej, jak na przykład Neo4j. Jednak w działających aktualnie rozwiązaniach e-Commerce bazy grafowe stanowią rzadkość. Obecnie głównym modelem wykorzystywanym w bazach danych jest model relacyjny (tab. 6.1).

**Tabela 6.1.** Ranking systemów bazodanowych sierpień 2021

| Pozycja w rankingu | | | **DBMS** | **Model Bazy** | Punkty | | |
|---|---|---|---|---|---|---|---|
| Aug 2021 | Jul 2021 | Aug 2020 | | | Aug 2021 | Jul 2021 | Aug 2020 |
| 1. | 1. | 1. | Oracle | Relational, Multi-model | 1269.26 | +6.59 | -85.90 |
| 2. | 2. | 2. | MySQL | Relational, Multi-model | 1238.22 | +9.84 | -23.36 |
| 3. | 3. | 3. | Microsoft SQL Server | Relational, Multi-model | 973.35 | -8.61 | -102.53 |
| 4. | 4. | 4. | PostgreSQL | Relational, Multi-model | 577.05 | -0.10 | +40.28 |
| 5. | 5. | 5. | MongoDB | Document, Multi-model | 496.54 | +0.38 | +52.98 |
| 6. | 6. | 7. | Redis | Key-value, Multi-model | 169.88 | +1.58 | +17.01 |
| 7. | 7. | 6. | IBM Db2 | Relational, Multi-model | 165.46 | +0.31 | +3.01 |
| 8. | 8. | 8. | Elasticsearch | Search engine, Multi-model | 157.08 | +1.32 | +4.76 |
| 9. | 9. | 9. | SQLite | Relational | 129.81 | -0.39 | +3.00 |
| 10. | 11. | 10. | Microsoft Access | Relational | 114.84 | +1.39 | -5.02 |
| 11. | 10. | 11. | Cassandra | Wide column | 113.66 | -0.35 | -6.18 |
| 12. | 12. | 12. | MariaDB | Relational, Multi-model | 98.98 | +0.99 | +8.06 |
| 13. | 13. | 13. | Splunk | Search engine | 90.60 | +0.55 | +0.69 |
| 14. | 14. | 15. | Hive | Relational | 83.93 | +1.26 | +8.64 |
| 15. | 15. | 17. | Microsoft Azure SQL | Relational, Multi-model | 75.15 | -0.06 | +18.31 |
| 16. | 16. | 16. | Amazon DynamoDB | Multi-model | 74.90 | -0.30 | +10.15 |
| 17. | 17. | 14. | Teradata | Relational, Multi-model | 68.82 | -0.13 | -7.96 |
| **18.** | **18.** | **21.** | **Neo4j** | **Graph** | **56.95** | **-0.21** | **+6.77** |
| 19. | 19. | 19. | SAP HANA | Relational, Multi-model | 55.57 | +1.76 | +2.46 |
| 20. | 20. | 20. | Solr | Search engine, Multi-model | 51.06 | -0.73 | -0.63 |

Wyniki rankingu prowadzonego przez DB-Engines pokazują różnicę pomiędzy najpopularniejszymi systemami relacyjnym (Oracle i MySQL) a najpopularniejszym systemem grafowym (Neo4j).

**Źródło:** (https://db-engines.com/en/ranking)

Grafy są strukturą abstrakcyjną, niemającą swojego bezpośredniego odzwierciedlenia w środowisku. Ich elementy, czyli wierzchołki i krawędzie, mogą mieć różne liczebności. Ponadto model matematyczny grafów nie jest w naturalny sposób odzwierciedlany w organizacji pamięci maszyn liczących (komputerów) (Horzyk). Implementując grafy, wykorzystuje się inne dobrze zdefiniowane struktury programistyczne, takie jak tablice i listy. Generalnie służą one do przechowywania informacji na temat sąsiednich (incydentnych) wierzchołków lub łączących je krawędzi (Goodrich, Tamassia 2001, rozdz. 6). Grafy reprezentujemy oraz implementujemy zwykle w postaci (Cormen 2009, rozdz. 6):

– Macierzy sąsiedztwa (*adjacency matrix*) – jest przedstawiana jako tablica dwuwymiarowa, gdzie indeksy wierszy i kolumn reprezentują numery wierzchołków, a wartości elementów równe 1 oznaczają krawędź łączącą wierzchołki określone numerem wiersza i kolumny.
– Macierzy incydencji (*incidence matrix*) – jest przedstawiana jako tablica dwuwymiarowa, gdzie indeksy wierszy reprezentują numery wierzchołków, a indeksy kolumn oznaczają numery krawędzi. Elementy równe 1 oznaczają krawędzie oznaczone numerami kolumn incydentnych z wierzchołkami oznaczonymi numerami wierszy.
– Listy sąsiedztwa (*adjacency list*) – jest przedstawiana jako lista, gdzie indeksy reprezentują numery wierzchołków, a każdy element tej listy jest listą numerów sąsiednich wierzchołków z numerem wierzchołka będącym w indeksie.

– Listy incydencji (*incidence list*) – jest przedstawiana jako lista, gdzie indeksy reprezentują numery wierzchołków, a każdy element tej listy jest listą numerów krawędzi incydentnych z numerem wierzchołka będącym w indeksie.
– Listy krawędzi (*list of edges*) – jest przedstawiana jako lista par numerów wierzchołków dla każdej krawędzi.

Tabela 6.2. Porównanie reprezentacji grafowych w pamięci komputerów

| Reprezentacja grafu | Złożoność pamięciowa |
|---|---|
| macierzy sąsiedztwa (*adjacency matrix*) | $O(|N|^2)$ |
| macierzy incydencji (*incidence matrix*) | $O(|N|*|E|)$ |
| listy sąsiedztwa (*adjacency list*) | $O(|N|+|E|)$ |
| listy incydencji (*incidence list*) | $O(|N|+|E|)$ |
| listy krawędzi (*list of edges*) | $O(|E|)$ |

Gdzie |N| to liczność zbioru wierzchołków, |E| to liczność zbioru krawędzi

**Źródło:** opracowanie własne

## Przykładowe systemy i algorytmy rekomendacji

### Algorytmy i systemy bazujące na grafach heterogenicznych

W wielu algorytmach i systemach rekomendacji rolę GRSs (*Graph-Based Recommender System*) stanowią grafy heterogeniczne. Generalnie pojęcie heterogeniczności jest związane z niejednorodnością (zróżnicowaniem) (*Słownik języka…* 2021). W obszarze grafów i sieci graf można nazwać heterogenicznym, jeśli zawiera różne typy węzłów i krawędzi (łuków) (Wang i in. 2019). W literaturze tego typu grafy określa się jako heterogeniczne sieci informacyjne *Heterogeneous Information Network*, określane skrótem HIN (Shi i in., 2015).

W GRSs podstawowymi typami węzłów (N) są: obiekty (*items*, $o \in O$) i użytkownicy (*users*, $u \in U$), natomiast typem krawędzi ($e \in E$) jest krawędź łącząca obiekt z użytkownikiem (($o, u) \in E$ ) (Hekmatfar, Haratizadeh, Goliaei 2021) (rys. 6.1). Jeśli do krawędzi zostaną dodane wagi na przykład reprezentujące oceny użytkowników nadawane obiektom, wówczas mamy do czynienia z grafem ważonym, który stanowi podstawę do konstruowania macierzy ocen (*rating matrix, RM*). Macierz tego typu stanowi podstawę dla całej grupy algorytmów rekomendacji typu Collaborative Filtering Methods. Między innymi są to takie algorytmy jak (*A Survey of Matrix…* 2018):
– ALS – *Alternative Least Square*;
– *Soft-Impute*;
– PFBS – *Proximal Forward-Backward Splitting*;
– ADDMM – *Alternating Direction Method Of Multipliers*;
– SVT – *Singular Value Thresholding*;

- APG – *Accelerated Proximal Gradient*;
- *Nonlinear Successive Over-Relaxation Algorithm*;
- SGD – *Stochastic Gradient Descent*;
- EM – *Expectation Maximization Algorithm*;
- *Funk* MF;
- SVD.

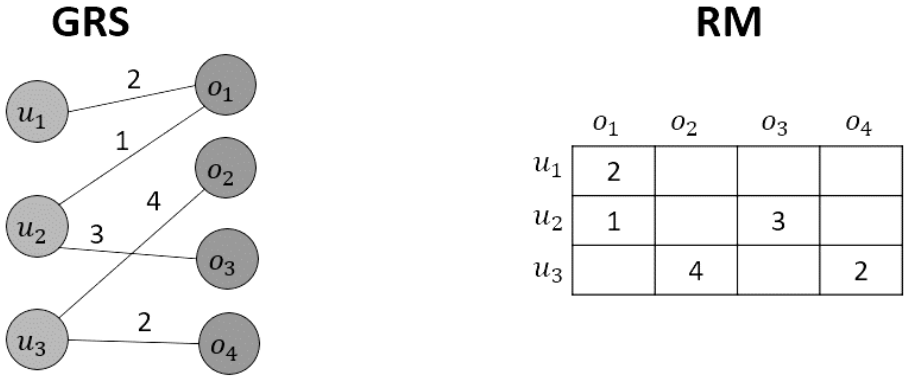

**Rysunek 6.1.** Przykład podstawowego modelu heterogenicznego GRS i jego odpowiednik w postaci macierzy ocen RM. Graf zawiera dwa typy węzłów powiązanych ze sobą krawędziami z wagami

**Źródło:** opracowanie własne

Kolejną grupę GRS stanowią tzw. grafy preferencji *Preference Graph Based Recommendation*, określane skrótem PGRec (Hekmatfar, Haratizadeh, Goliaei 2021). Główną ideą PGRec jest modelowanie problemu systemu rekomendującego jako problemu przewidywania wagi w nowym typie grafu heterogenicznego. Graf tego typu posiada dodatkowy typ węzłów w postaci preferencji (*preferences*, $p \in P$). Inna nazwa spotykana w literaturze dla tego typu grafów to *Tripartite Preference Graph*, czyli TPG (Shams, Haratizadeh 2017).

PGRec jest heterogenicznym grafem ważonym, który w swojej najprostszej postaci jest grafem trójdzielnym, w którym $N = U \cup P \cup O$ i $E = E_{UO} \cup E_{PO} \cup E_{UP}$ (rys. 6.2). $E_{UO}$ jest zbiorem ważonych krawędzi między użytkownikami a obiektami opartymi na macierzy ocen (**RM**). $E_{PO}$ reprezentuje ważone krawędzie łączące węzeł preferencji z węzłem obiektu o wadze $W_{PU} \in \{-1,1\}$ ($w_{p_{ij}o} = 1$ i $w_{p_{ij}o} = -1$). $E_{UP}$ to zestaw ważonych krawędzi łączących węzeł użytkownika z węzłem preferencji z wagą $W_{UP} \in [-(k_{max} - k_{min}), (k_{max} - k_{min})]$. Z kolei $w_{up_{ij}} = r_{ui} - r_{uj}$ mówi nam, jak bardzo użytkownik *u* preferuje obiekt *i* nad obiekt *j*. Ponadto każdy użytkownik *u* wykazuje zainteresowanie każdym obiektem *o* z oceną $r_{uo}$. Może być ona określona liczbowo {*1, 2, 3, 4, 5*} lub binarnie {*0* jako niechęć, *1* jako polubienie}). Istnieje tylko jedna ocena użytkownika *u* dla obiektu *o*.

Zastosowanie grafów i sieci w systemach rekomendacji

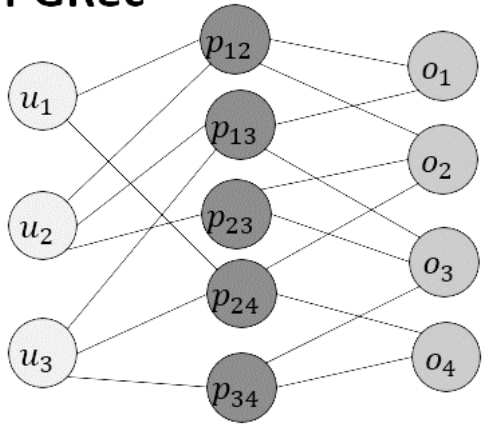

**Rysunek 6.2.** Przykład PGRec. Graf zawiera trzy typy węzłów.
Nowy typ węzłów stanowią preferencje

**Źródło:** opracowanie własne

Bardziej rozbudowana forma PGRec zawiera dodatkowe informacje o użytkownikach (grupy, $g \in G$) i obiektach (kategorie, $k \in K$) (rys. 6.3). Jest to realizowane poprzez dodanie kolejnych typów węzłów do grafu. Takie węzły są połączone krawędziami nieważonymi z odpowiednim użytkownikiem lub obiektami. Ponadto można dodać krawędzie w ramach poszczególnych warstw, np. pomiędzy użytkownikami (*U-U*) lub obiektami (*O-O*).

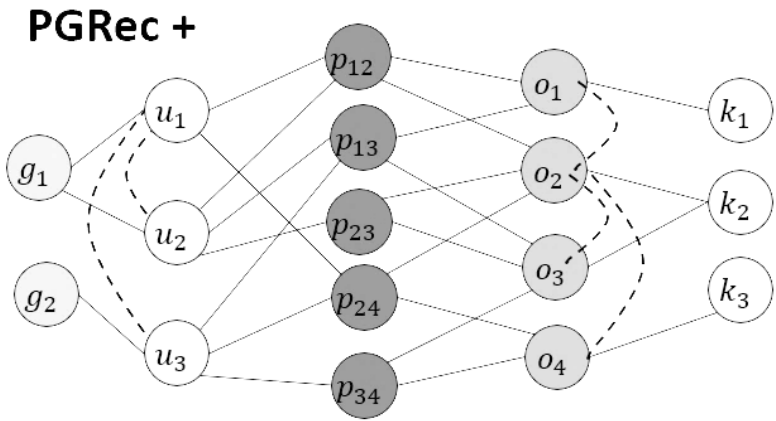

**Rysunek 6.3.** Przykład rozbudowanego PGRec. Graf zawiera 7 typów węzłów.
W stosunku do pierwotnego modelu PGRec zostały dodane typy węzłów zawierające
dodatkowe informacje o użytkownikach i obiektach

**Źródło:** opracowanie własne

Systemy rekomendacji wykorzystujące opisany powyżej graf preferencji to przede wszystkim systemy bazujące na rodzinie algorytmów GRank (Shams, Haratizadeh 2017). W ostatnich latach obserwuje się gwałtowny wzrost badań nad HIN. Jedną z metod badania tych struktur jest analiza ścieżek i ich semantyczna interpretacja (tab. 6.3), która również znajduje zastosowanie w budowie algorytmów rekomendacji (Shi i in. 2015).

**Tabela 6.3.** Przykłady ścieżek i ich interpretacja w odniesieniu do technik rekomendacji

| Wzór ścieżki | Semantyczna interpretacja | Technika rekomendacji |
|---|---|---|
| *UU* | Znajomi | *Social recommendation* |
| *UGU* | Użytkownicy w tej samej grupie (z tą sama cechą) | *Member recommendation* |
| *UOU* | Użytkownicy, którzy interesują się tymi samym produktem | *Collaborative recommendation* |
| *UOKOU* | Użytkownicy, którzy interesują się produktami tej samej kategorii (z tą samą cechą) | *Content recommendation* |

Do budowy wzorów ścieżek zostały wykorzystane typy (zbiory) węzłów występujące w grafie PGRec+ będącym również przykładem HIN

**Źródło:** opracowanie własne

## Algorytm rekomendacji bazujący na sesjach rekomendacji (ARS)

Algorytm ten ma na celu rozwiązanie zadania szczególnego (2) i można zaliczyć go do typu algorytmów hybrydowych, ponieważ może on dokonywać rekomendacji, opierając się zarówno na technice *Content Based*, jak i *Collaborative Filtering* (Malinowski 2020).

W rozwiązaniu zaproponowano budowę modelu danych opartą na bazie pojęcia „sesji rekomendacji". W obszarze aparatu matematycznego pojęcie to jest zdefiniowane na podstawie dwudzielnego unigrafu skierowanego $G$ (Wojciechowski 2019), zwanego grafem sesji rekomendacji takiego, że:

$$G = <N, E>n \qquad (3)$$

gdzie:
$N = J \cup O$ – zbiór wierzchołków (*node*)
$E \subset J \times O$ – zbiór łuków (krawędzi skierowanych) (*directed edge*)
$O$ – zbiór obiektów, gdzie **obiekt** (*object*) ($o \in O$) może stanowić:
- towar w sklepie internetowym;
- film w wypożyczalni na żądanie;
- pracownik w serwisie związanym z zatrudnieniem;
- artykuł prasowy w serwisie informacyjnym;
- osoba w serwisie społecznościowym.

$J$ – zbiór jąder, gdzie **jądro** (*kernel*) ($j \in J$) może stanowić:
- kategoria produktów – jeden z podzbiorów produktów mających wspólne cechy;

- zamówienie – wynik działań klienta w sklepie zakończony zakupem;
- lista życzeń klienta – podzbiór produktów sklepu związanych z klientem, wynikający z jego chęci zakupowych;
- ekspert – podzbiór produktów wskazanych przez specjalistę dziedzinowego;
- identyfikator odwiedzin strony WWW – unikalny klucz nadawany odwiedzinom użytkownika w serwisie WWW – odwiedziny składają się z ciągu obejrzanych stron WWW sklepu internetowego;
- osoba – struktura informatyczna identyfikująca i opisująca użytkownika w systemie informatycznym.

Ze względu na przyjętą i omówioną w dalszej części interpretację modelu i jego składowych zakłada się spełnienie następujących ograniczeń (rys. 6.4):
- $J \cap O = \emptyset$ – żadne jądro nie może być obiektem i żaden obiekt nie może być jądrem;
- $(\forall j \in J)(\exists o \in O)(\exists e \in E)(e = (j, o))$ – każde jądro musi być związane z co najmniej jednym obiektem;
- $(\forall o \in O)(\exists j \in J)(\exists e \in E)(e = (j, o))$ – każdy obiekt musi być związany z co najmniej jednym jądrem.

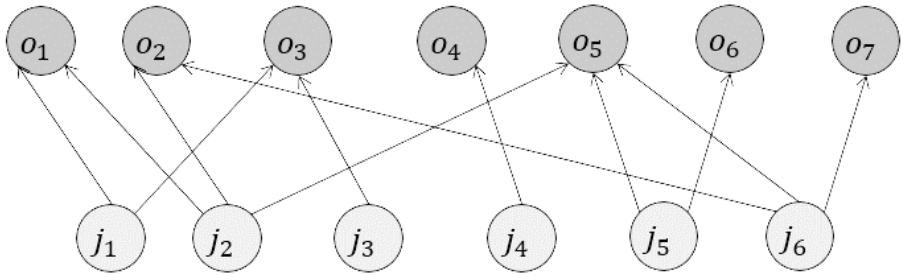

**Rysunek 6.4.** Graf sesji rekomendacji złożony z obiektów {o_1, o_2, o_3….o_n} , jąder {j_1, j_2, j_3….j_n}

**Źródło:** opracowanie własne

Pojedynczą sesję $S$ można przedstawić jako podgraf grafu $G$, taki że:
$$S = <N', E'> \qquad (4)$$
gdzie:
$N' = j \cup O'$ – zbiór wierzchołków sesji,
$E' \subset E$ – zbiór łuków sesji,
$O' \subset O$ – zbiór obiektów sesji (związanych z jądrem),
$j \in J$ – jądro sesji.

Zakłada się spełnienie następujących ograniczeń (rys. 6.5):
- $(\forall o \in O')(\exists e \in E)(e = (j, o))$ – każdy obiekt sesji powiązany jest z jądrem;
- $(\nexists o \in O \setminus O')(\exists e \in E)(e = (j, o))$ – nie istnieje obiekt powiązany z jądrem niebędący elementem sesji.

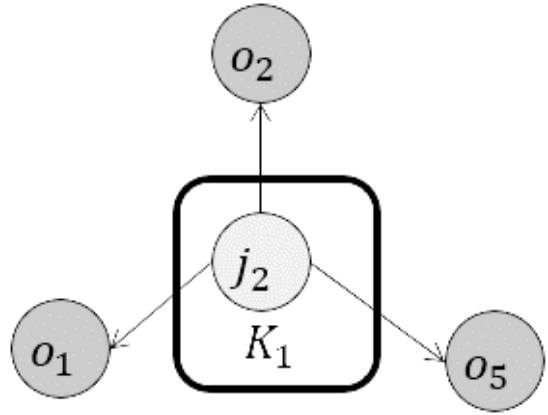

**Rysunek 6.5.** Model pojedynczej sesji rekomendacji złożony z obiektów {o_1, o_2, o_3} , jądra { j_2} oraz klasy { K_1}

**Źródło:** opracowanie własne

Sesja złożona jest tylko z jednego jądra i wszystkich łuków oraz obiektów z nim związanych.

$S$ – zbiór sesji, gdzie **sesję** ($s \in S$) może stanowić:
– kategoria produktów wraz z jej produktami;
– zamówienie wraz z pozycjami;
– lista życzeń klienta wraz z elementami;
– ekspert wraz ze swoimi ocenami;
– identyfikator odwiedzin strony WWW wraz z odwiedzonymi stronami;
– osoba wraz ze znajomymi.

Ponadto, ze względu na swoje właściwości fizyczne i podobieństwa, oznaczmy przez $K$ zbiór różnych typów jąder sesji. Każdemu $k \in K$ odpowiada $J_k \subset J$ podzbiór jąder tego samego typu, czyli o tych samych właściwościach funkcjonalnych np.: kategorie produktów, zamówienia itd.

Niech $K \subset J$ – klasa jest podzbiorem zbioru jąder, gdzie **klasę** ($K$) jąder sesji mogą stanowić:
– kategorie produktów;
– zamówienia;
– listy życzeń klientów;
– eksperci;
– identyfikatory odwiedzin serwisu WWW;
– osoby.

Ponadto zakłada się spełnienie następujących ograniczeń:
– $(\forall i \neq j)(K_i \cap K_j = \emptyset)$ – żadne jądro nie występuje w dwóch i więcej klasach;
– $(\forall j \in J)(\exists i)(j \in K_i)$ – każde jądro należy do jakiejś klasy.

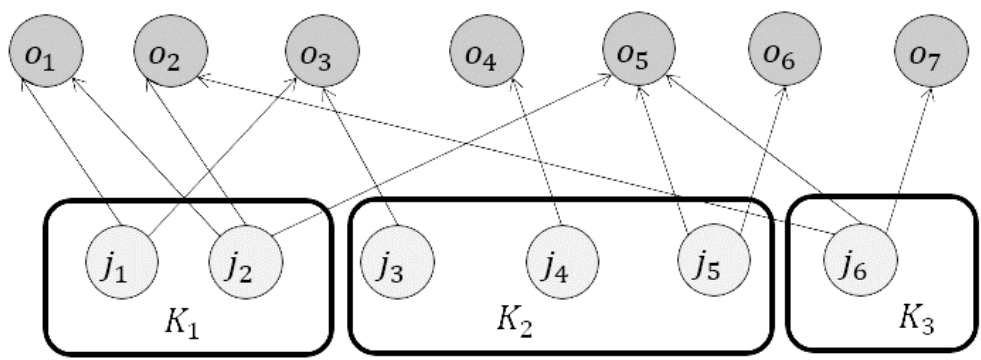

**Rysunek 6.6.** Klasy jąder {K_1, K_2, K_3} złożone z jąder {j_1, j_2, j_3….j_n}.

**Źródło:** opracowanie własne

Klasy jąder można podzielić na następujące typy:
- Behawioralne – jądra powstające w wyniku działań użytkowników. Są one bardzo zmienne w czasie, jak np.: odwiedziny, zakupy lub lista życzeń.
- Statyczne – jądra związane z cechami obiektów. Są one niezmienne w czasie, jak np.: rozmiar, kolor lub przynależność do kategorii.
- Mieszane – jądra powstające w wyniku oddziaływania otoczenia systemu, ale słabo zmienne w czasie, jak np.: wskazania ekspertów lub zewnętrzne rankingi.

Algorytm ARS składa się z określonego typu danych wejściowych, wyjściowych i kroków. Szczegółowo są one omówione wraz z przykładami w artykule *Recommendation Algorithm Based on Recommendation Sessions* zawartym w materiałach pokonferencyjnych *36th International Business Information Management Association (IBIMA)* (Malinowski 2020).

**Dane wejściowe:**
$G$ – graf sesji;
$m$ – obiekt (węzeł grafu), do którego mają zostać dowiązane rekomendacje.

**Dane wyjściowe:**
$R_m$ – wektor rekomendacji dla obiektu $m$ (im niższa pozycja w wektorze, tym lepsza rekomendacja dla obiektu $m$).

Formalnie zależność pomiędzy zbiorem $R_m^*$ (2) a wektorem rekomendacji $R_m$ jest taka, że każdy element zbioru rekomendacji $R_m^*$ jest elementem wektora rekomendacji, ale nie każdy element wektora $R_m$ jest elementem zbioru $R_m^*$. Wynika to z faktu, że algorytm w wyniku nie zwraca tylko i wyłącznie najlepszego dopasowania, ale również zbliżone do najlepszych $N$ obiektów (tak zwane „Top-N").

**Kroki:**

*S01*: Podanie danych wejściowych

*S02:* Budowa podgrafu $G'_m$ grafu $G$ złożonego z węzła $m$ i węzłów sąsiadujących z węzłem $m$ oraz łuków pomiędzy nimi a węzłem $m$

*S03*: Budowa podgrafu $G''_m$ grafu $G$ złożonego z grafu $G'_m$ i węzłów sąsiadujących z węzłami grafu $G''_m$ oraz łuków pomiędzy nimi a węzłami grafu $G'_m$

*S04:* Oszacowanie stopni wchodzących dla każdego węzła będącego obiektem podgrafu $G''_m$

*S05:* Posortowanie malejąco obiektów (węzłów) względem stopnia wchodzącego

*S06*: Zapisanie w wektorze $R_m$ posortowanych obiektów bez obiektu $m$

Do reprezentacji grafu $G$ w docelowym systemie e-Commerce została wybrana reprezentacja w postaci listy krawędzi.

Implementacja algorytmu rekomendacji ARS została przeprowadzona w rozwiązaniu e-Commerce w postaci platformy sklepu internetowego z grami planszowymi, funkcjonującym pod adresem elektronicznym (URL): https://am76.pl, zwanym AM76. Jako metoda implementacji algorytmu przyjęty został standard SQL zaimplementowany w relacyjnej bazie danych systemu informatycznego AM76. Algorytm znalazł również zastosowanie w systemie rekomendacji funkcjonującym w *Encyklopedii Gier Planszowych* dostępnej pod adresem elektronicznym (URL) http://gra24h.pl (Malinowski, 2021).

## *Page Rank*, czyli tajemnica Google

*Page Rank* pojawił się w 1998 roku, został stworzony przez założycieli Google – Larry'ego Page i Sergeya Brina. Jest on wyliczany automatycznie przez opatentowany algorytm i ma za zadanie wiarygodnie oceniać strony internetowe (Brin, Page 1998). *Page Rank*, czyli ranking stron WWW, to ocena wartości strony wyrażana w postaci liczby na podstawie jakości oraz ilości prowadzących do witryny linków z innych stron.

W swej podstawowej wersji jest on wykorzystywany przez wyszukiwarkę internetową Google, która nadaje stronom WWW określone wagi, które zależą od liczby innych stron, które na nią wskazują. Wartość wag można wykorzystać do uszeregowania wyników zapytań. Ten ranking stron byłby jednak mało odporny na zjawisko znane jako spam, ponieważ dość łatwo jest sztucznie tworzyć wiele stron wskazujących na istotną dla nas stronę (Arasu i in. 2001). Aby przeciwdziałać takim praktykom, algorytm *Page Rank* rozszerza podstawową ideę linków, biorąc pod uwagę znaczenie każdej strony, która wskazują na analizowaną stronę. Oznacza to, że definicja wag stron (*Page Rank*) jest „cykliczna", czyli waga strony zależy od wag stron wskazujących na nią i jednocześnie wpływa na ważność stron, na które wskazuje. Model matematyczny, na którym bazuje algorytm, oparty jest o teorię grafów i sieci. W podstawowej wersji algorytmu *Page Rank* definiuje się graf $G$, taki że:

$$G = <N, E>  \quad (5)$$

gdzie:

$N$ — zbiór wierzchołków (*node*), gdzie wierzchołek ($n \in N$) to strona WWW

$E \subset N \times N$ – zbiór łuków (krawędzi skierowanych) (*directed edge*), gdzie łuk ($e \in E$) to link pomiędzy stronami WWW.

Ponadto:

$in(n)$ – zbiór wszystkich stron linkujących do strony $n$,

$out(n)$ – zbiór wszystkich stron linkujących od strony $n$,

$PR(n)$ – *Page Rank* strony $n \in N$,

$d$ – współczynnikiem normalizującym.

Podstawowy wzór rankingu wygląda tak:

$$PR(u) = d \sum_{v \in in(u)} \frac{PR(v)}{|out(v)|} \quad (6)$$

Dla współczynnika normalizującego, zwanego również współczynnikiem tłumienia, autorzy algorytmu przyjęli wartość $d = 0{,}85$ (Page, Brin 1999). Całe równanie jest rekurencyjne. Oryginalna hipoteza autorów algorytmu zakłada, że średnia z wszystkich wartości *Page Rank* wszystkich witryn wynosi 1.

### Kroki podstawowej wersji algorytmu:

*S01:* Zainicjuj wszystkie rangi na 1/(liczba wszystkie strony).

$$\forall n \in N \ \ PR(n) = \frac{1}{|N|} \quad (7)$$

*S02 do Sstop*: Dla każdej strony $u$ zaktualizuj jej ranking, aby był iloczynem $d$ i sumy rankingów stron linkujących $v$ z wcześniejszej iteracji podzielony przez liczbę linków ze strony $v$ zgodnie ze wzorem (6)

− *Sstop:* Stop algorytmu. Powinien nastąpić, gdy *Page Rank* ustabilizuje się lub zbiegnie, czyli:
− *Page Rank* każdej strony z poprzedniej iteracji różni się od bieżącej iteracji o mniejszą lub równą marginesowi błędu (na przykład 0,01);
− lub po wykonaniu 100 iteracji
  w zależności od tego, co nastąpi wcześniej.

### Przykład użycia:

Dla przedstawionego grafu $G$ (rys. 6.8) został zbudowany ranking stron. Dla uproszczenia obliczeń zostało przyjęte $d = 1$.

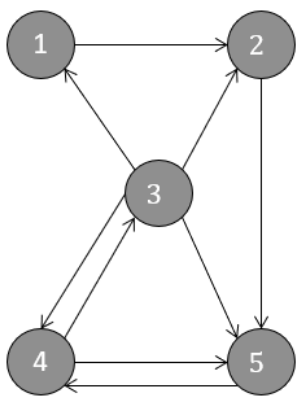

**Rysunek 6.6.** Graf G reprezentujący linki pomiędzy stronami.
Struktura grafu odzwierciedla 5 stron WWW powiązanych wzajemnymi linkami

**Źródło:** opracowanie własne

Poszczególne kroki algorytmu dają następujące wyniki przedstawione w tabeli 6.4.

**Tabela 6.4.** Wyniki działania podstawowej wersji algorytmu Page Rank

|        | *S01* | *S02* | *S03* | Ranking |
|--------|-------|-------|-------|---------|
| **PR(1)** | 1/5 | 1/20 | 1/40 | 5 |
| **PR(2)** | 1/5 | 5/20 | 3/40 | 4 |
| **PR(3)** | 1/5 | 2/20 | 5/40 | 3 |
| **PR(4)** | 1/5 | 5/20 | 15/40 | 2 |
| **PR(5)** | 1/5 | 7/20 | 16/40 | 1 |

Po 3 iteracjach algorytmu został oszacowany ranking stron w oparciu o wartości *PR* dla każdej ze stron.

**Źródło:** opracowanie własne

### Nowa postać algorytmu *Page Rank*

Aktualnie podstawowy wzór wykorzystywany w szacowaniu *Page Rank* ma postać:

$$PR(u) = \frac{1-d}{|N|} + d \sum_{v \in in(u)} \frac{PR(v)}{|out(v)|} \qquad (8)$$

gdzie opis poszczególnych składowych wzoru jest analogiczny do wzorów (5) i (6)

### Szacowanie *Page Rank* w oparciu o macierz

Skutecznym, praktycznym sposobem na znalezienie wektora *Page Rank* **r** jest użycie języka i metody algebry liniowej. Używając algebry liniowej, wektor *Page Rank* **r** może być oszacowany poprzez rozwiązanie jednorodnego układu liniowego (Worwa, Konopacki 2013):

$$(A^T - I)r^T = 0^T \qquad (9)$$

lub rozwiązując problem z wektorami własnymi:

$$r = r \cdot A \qquad (10)$$

gdzie:
$r^T$ – jest wektorem kolumnowym transponowanym do wektora wierszowego $r$,
$I$ – jest macierzą jednostkową rzędu m,
$0^T$ – jest wektorem kolumnowym wszystkich zer,
$m = |N|$ – jest liczbą stron WWW,
$A^T$ – jest transponowaną macierzą $A = [a_{ij}]_{mxm}$, której elementy $a_{ij}$ są zdefiniowane w następujący sposób (rys. 6.6):

$$a_{ij} = \begin{cases} \frac{1}{IN(i)} & \text{jeśli strona } i \text{ jest zlinkowana ze stroną } j \\ 0 & \text{w innym przypadku} \end{cases} \qquad (11)$$

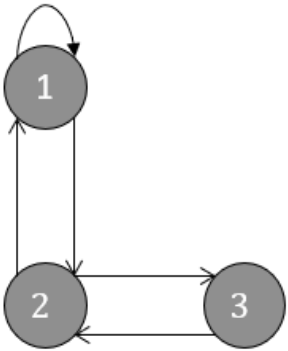

**Rysunek 6.6.** Odwzorowanie macierzy A dla grafu G. Pola macierzy macierz A zostały oszacowane zgodnie ze wzorem (11)

**Źródło:** opracowanie własne

Ponadto należy uzupełnić układ o równanie normalizacyjne w postaci:

$$r \cdot 1^T = 1 \qquad (12)$$

gdzie:
$1^T$ – jest wektorem kolumnowym o elementach 1.

Aktualnie większość współczesnych wyszukiwarek internetowych wykorzystuje w swoich silnikach wyszukiwawczych jednej z form algorytmu *Page Rank*.

## Podsumowanie

Teoria grafów i sieci znajduje zastosowanie w obszarze rekomendacji przede wszystkim na poziomie modeli matematycznych algorytmów rekomendacji, na podstawie których budowane są systemy rekomendacji implementowane w systemach informatycznych. W modelach tych wykorzystywane są głównie grafy heterogoniczne z dwoma podstawowymi typami węzłów w postaci obiektów i użytkowników, które mogą być rozbudowywane o inne typy w zależności od przyjętego modelu.

Znacznie wyższy poziom wykorzystania relacyjnych baz danych w stosunku do baz grafowych w obszarze przetwarzania danych powoduje, że fizycznie wykorzystywane są ich reprezentacje w postaci macierzy i list, które są możliwe do implementacji w relacyjnych bazach danych. Trend ten wraz z rozwojem baz grafowych, takich jak np. Neo4j, będzie się najprawdopodobniej zmieniał, co w efekcie pozwoli na efektywne wykorzystanie grafów i sieci, szczególnie na poziomie implementacji systemów rekomendacji oraz ich bieżącego funkcjonowania.

## Literatura

# APPLICATION OF GRAPHS AND NETWORKS IN RECOMMENDATION SYSTEMS

**Abstract:** The aim of this chapter is to present the use of graph and network theory in the area of recommendations, in particular mathematical models constituting the basis for algorithms and recommendation systems constructed on their basis. The first part of the chapter contains a synthetic description of the recommendation area with particular emphasis on the types of recommendation solutions and a mathematical description of the problem. The next part of the chapter presents models and techniques for the use of graphs and networks, as well as sample algorithms based on them.

**Keywords:** recommendation system, recommendation algorithm, graphs and networks